\documentclass[conference,a4paper]{IEEEtran}
\IEEEoverridecommandlockouts
\usepackage{cite}
\usepackage{indentfirst}
\usepackage{multirow}
\usepackage{booktabs}
\usepackage{makecell}
\usepackage{float}
\usepackage{amsmath,amssymb,amsfonts}
\usepackage{algorithmic}
\usepackage{graphicx}
\usepackage{graphicx,subfigure}
\graphicspath{{./figures/}}
\usepackage{textcomp}
\usepackage{xcolor}
\usepackage{subfigure}
\usepackage{booktabs}
\usepackage{bm}
\usepackage{amsthm}
\usepackage{epstopdf}
\def\BibTeX{{\rm B\kern-.05em{\sc i\kern-.025em b}\kern-.08em
		T\kern-.1667em\lower.7ex\hbox{E}\kern-.125emX}}

\begin{document}
	\bibstyle{IEEEtran}
	\title{6G LLM Agents: A Novel Paradigm for Task-Oriented Physical-Layer Automation}

	\author{
		\IEEEauthorblockN{Zhuoran~Xiao,
			Chenhui~Ye$\IEEEauthorrefmark{2}$,
			Yunbo~Hu,
			Honggang~Yuan,
			Yihang~Huang,
			Yijia~Feng,
			Liyu~Cai,
			and Jiang~Chang
}

		\IEEEauthorblockA{
			Nokia Bell Labs, Shanghai, China\\
			E-mails: \{zhuoran.xiao, chenhui.a.ye$\IEEEauthorrefmark{2}$, yunbo.hu, honggang.yuan, yihang.huang, yijia.feng, liyu.cai, jiang.chang\}\\@nokia-sbell.com
		}
	}
\maketitle

\begin{abstract}
The rapid advancement in generative pre-training models is propelling a paradigm shift in technological progression from basic applications such as chatbots towards more sophisticated agent-based systems. It is with huge potential and necessity that the 6G system be combined with the copilot of large language model (LLM) agents to manage the highly complicated communication system with new emerging features such as native AI service and sensing. With the 6G-oriented agent, the base station could understand the transmission requirements of various dynamic upper-layer tasks, automatically orchestrate the optimal system workflow, and raise the performance accordingly. Differing from existing LLM agents designed for general application, the 6G-oriented agent aims to make highly rigorous and precise planning with a vast amount of extra expert knowledge, which inevitably requires a specific system design from model training to implementation. This paper proposes a novel comprehensive approach for building task-oriented 6G LLM agents. We first propose a two-stage continual pre-training and fine-tuning scheme to build the field basic model and diversities of specialized expert models for meeting the requirements of various tasks. Further, a novel retrieval-based agent inference framework for leveraging the existing communication-related functions is proposed. Experiment results of exemplary tasks, such as 3GPP protocol Q\&A and physical-layer task decomposition, show the proposed paradigm's feasibility and effectiveness.
\end{abstract}


\section{Introduction} \label{intro}
The wireless communication system is currently in a new era of evolving from the traditional one designed solely for symbol transmition to a more functional and complicated one. The introduction of AI (Artificial Intelligence) methods into the air interface empowers the system to achieve enhanced performance across a variety of scenarios \cite{10360202}, and the new features such as JCAS (Joint Communication and Sensing) greatly extend the functions of the system \cite{10012421}. Moreover, there emerges novel application scenarios, such as autonomous driving \cite{10437455} and augmented reality \cite{10322917}, which introduce diverse requirements beyond traditional use cases. In addition, with the goal of optimizing the system specifically for a certain kind of downstream task, task-oriented communication is deemed a promising approach to maximize overall system efficiency \cite{10283741,9606667}. Relevant research has demonstrated that through specialized system designs, the performance can be significantly boosted. It is evident that all the aforementioned necessitates the system's capability to automatically adapt to diverse communication scenarios and respond to various UE tasks. In the present communication system, the base station is restricted to working with predefined workflows and lacks comprehensive task understanding, causing the system to work inefficiently. This inefficiency not only squanders spectrum resources and energy but also negatively impacts the user experiences.

\begin{figure*}[htb!]
	\centering
	\includegraphics[width=0.7\textwidth]{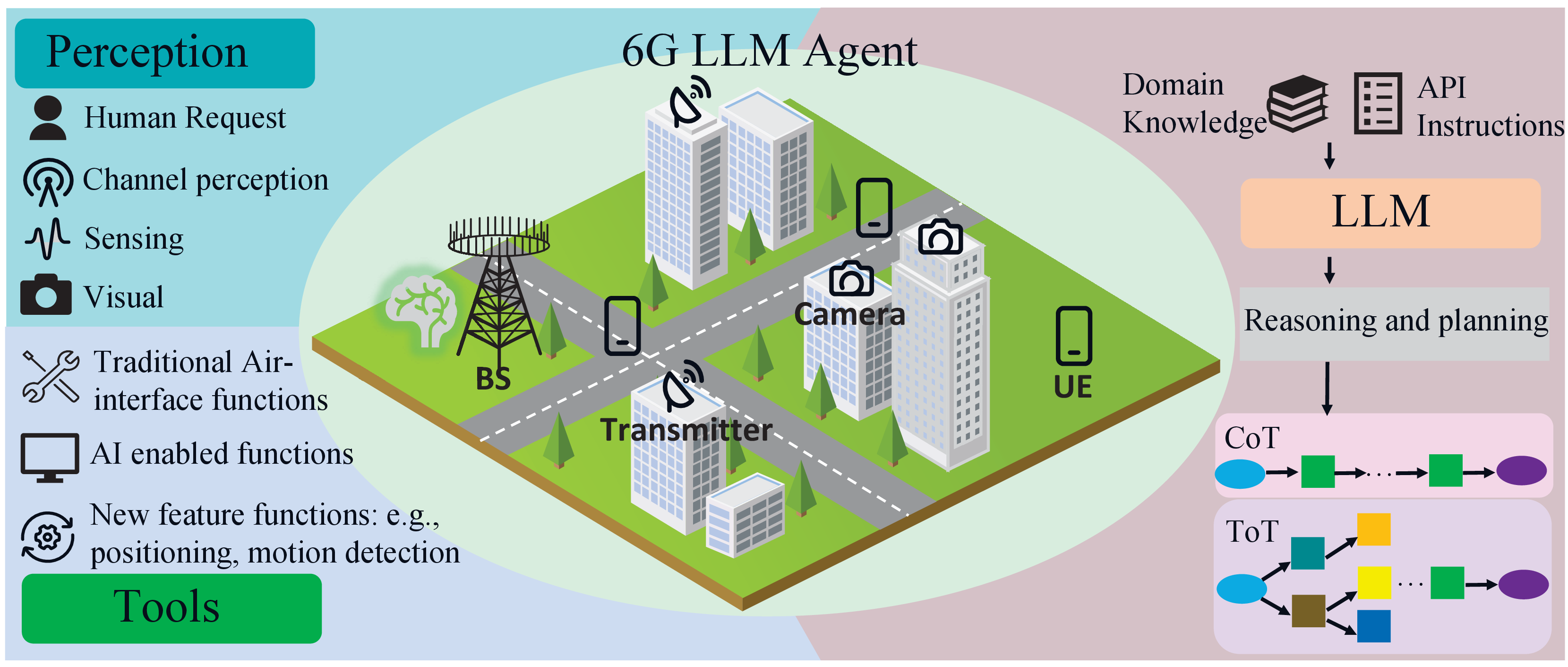}
	\caption{System framework of the proposed 6G LLM agent.}
	\label{system_fig}
	\vspace{-.5cm}
\end{figure*}

Through leveraging the powerful learning ability of large language models (LLMs), AI agents have demonstrated significant potential in orchestrating complex systems and formulating sequential plans and decisions \cite{10435998,10378628}. For a comprehensive AI agent, there are several essential functions, including observing the environment, formulating plans and making decisions, utilizing tools, and self-improvement through reinforcement learning. It is obvious that the vision of future wireless communication systems impeccably aligns with the prerequisites for developing LLM agents. First of all, the communication system possesses a solid ability to observe the scattering environments through multi-modal perception, including channel estimation, sensing, and imaging via electromagnetic waves, as well as visual capture through cameras. Secondly, there is a vast repository of accessible domain-specific data that includes protocol documents, software code repositories, and error logs, which can be utilized to train an LLM. 
This training would empower the system to perform planning and make decisions, such as devising system workflows and configuring system settings. Thirdly, a multitude of functional APIs (Application Programming Interface) already exist in the communication system, which can be regarded as the toolkit for agents, facilitating various operations. Finally, performance metrics, such as the QoS (quality of service) and QoT (quality of task), can serve as reinforcement rewards to refine the agent's actions.

Owing to the distinct and unique characteristics, the 6G-oriented LLM agent exhibits marked differences from general-purpose LLMs. For instance, an advanced 6G LLM agent necessitates extensive learning from a vast corpus of domain-specific knowledge. Besides, utilizing tools within most existing agent systems is trivial due to the limited variety of tools available. In contrast, for a sophisticated wireless communication system, many API tools with complicated invoking logic exist, thereby presenting substantial challenges for effective implementation and management.

In order to address the issue mentioned above, we propose an innovative system paradigm for the training and inferencing process of the task-oriented 6G LLM agents. Specifically, we propose a two-stage continual training methodology. In the first stage, we aim to inject domain-specific knowledge into a pre-training general LLM, enhancing its capacity to understand the communication tasks and orchestrate the system. The resulting model from this phase is termed the field basic model since it can be further fine-tuned to adapt to specified communication scenarios or tasks. In the second stage, we use a small amount of high-quality data to fine-tune the level-$1$ model. This refinement ensures that the model's planning and decision-making capabilities are optimally tailored to the inherent functionalities of the host system. The adaptability of this fine-tuning stage allows for easy adoption across diverse scenarios and applications. Moreover, we devise a communication system-specified retrieval approach to broaden the agent's proficiency in utilizing tools. This method significantly improves the accuracy of invoking professional functions compared to existing methodologies.

The remainder of this paper is organized as follows. The system framework is described in section \ref{system}. The two-stage LLM training approach is given in section \ref{2-stage}. Section \ref{agent} introduces the proposed workflow of the LLM agents. Section \ref{experiment} introduces our experiment setup and corresponding results, which evaluates the performance of our proposed approach from different perspectives. Section \ref{conclusion} draws our main conclusions.

\begin{figure}
	\centering
	\includegraphics[width=0.45\textwidth]{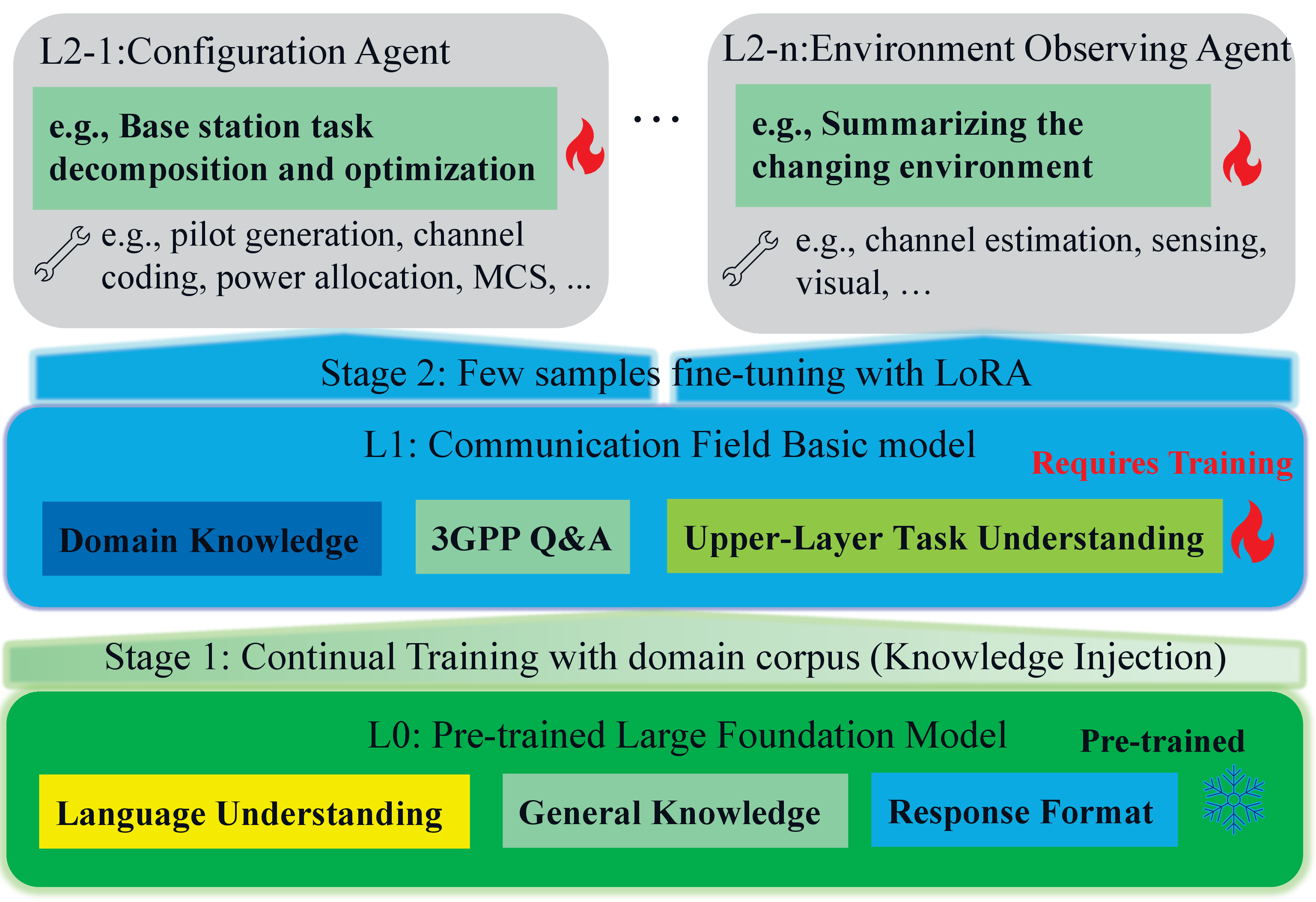}
	\caption{The proposed three model levels and the corresponding functions. The L1 model is trained from the pre-training L0 model and serves as the domain basic model. L2 models are fine-tuned based on the L1 model tailored for specific target scenarios or agent roles.}
	\label{3-layers}
	\vspace{-.5cm}
\end{figure}

\section{System Framework} \label{system}
We consider a typical scenario in which the LLM agent is deployed at the base station to orchestrate the entire radio access network (RAN) within a cell. As shown in Fig. \ref{system_fig}, the 6G LLM agents system is mainly composed of three parts: the LLM, multi-source perception, and tools. The LLM serves as the brain of the system. It first takes the upper-layer task description and the environment perception as the input, thus understanding the transmission requirements. Then, it outputs the reasoning and planning via the chain of thoughts (CoT) or tree of thoughts (ToT), thereby dynamically reconfiguring the system's setup and workflow accordingly.

The distinguishing factors that set the 6G agent apart from other applications lie in its advanced capacity to perceive the scattering environment and adeptly utilize tools. This perception capability is significantly enhanced by integrating multi-modal abilities such as channel estimation, sensing, vision, and digital twins. With an increasingly precise understanding of the environment, the system is enabled to make effective adjustments. Moreover, future communication systems are poised to integrate more comprehensive tool sets, featuring cutting-edge functionalities like massive multiple-input multiple-output (MIMO) technology and AI-empowered physical layer functions, thereby enhancing their overall performance and versatility. In totality, these interconnected modules collectively constitute the architecture of the highly sophisticated 6G agent.

\section{Two-Stage Model training} \label{2-stage}
Due to the complexity of the natural environment and the diverse roles of the devices being applied, it is not practical to deploy a universal model for any circumstances. Thus, we propose a three-level model architecture shown in Fig. \ref{3-layers}. This framework commences with a foundational pre-training model, referred to as the L0 model, which is designed for general-purpose targets such as understanding natural language and formatting the outputs. Building upon this foundation model, we develop an L1 model through knowledge injection. This L1, or field basic model, encapsulates a reservoir of common domain-specific knowledge and earns expert ability for solving domain-related problems. Subsequently, by harmonizing the generalized knowledge embodied in the L1 model with the unique circumstances encountered by a particular deployment device, we derive the L2 model, which helps reasoning and planning and can be directly deployed in the practical 6G system.
\begin{figure}
	\centering
	\subfigure[Stage 1: Continual Training]{\label{fpt}
		\begin{minipage}[t]{0.7\linewidth}
			\centering
			\includegraphics[width=1\textwidth]{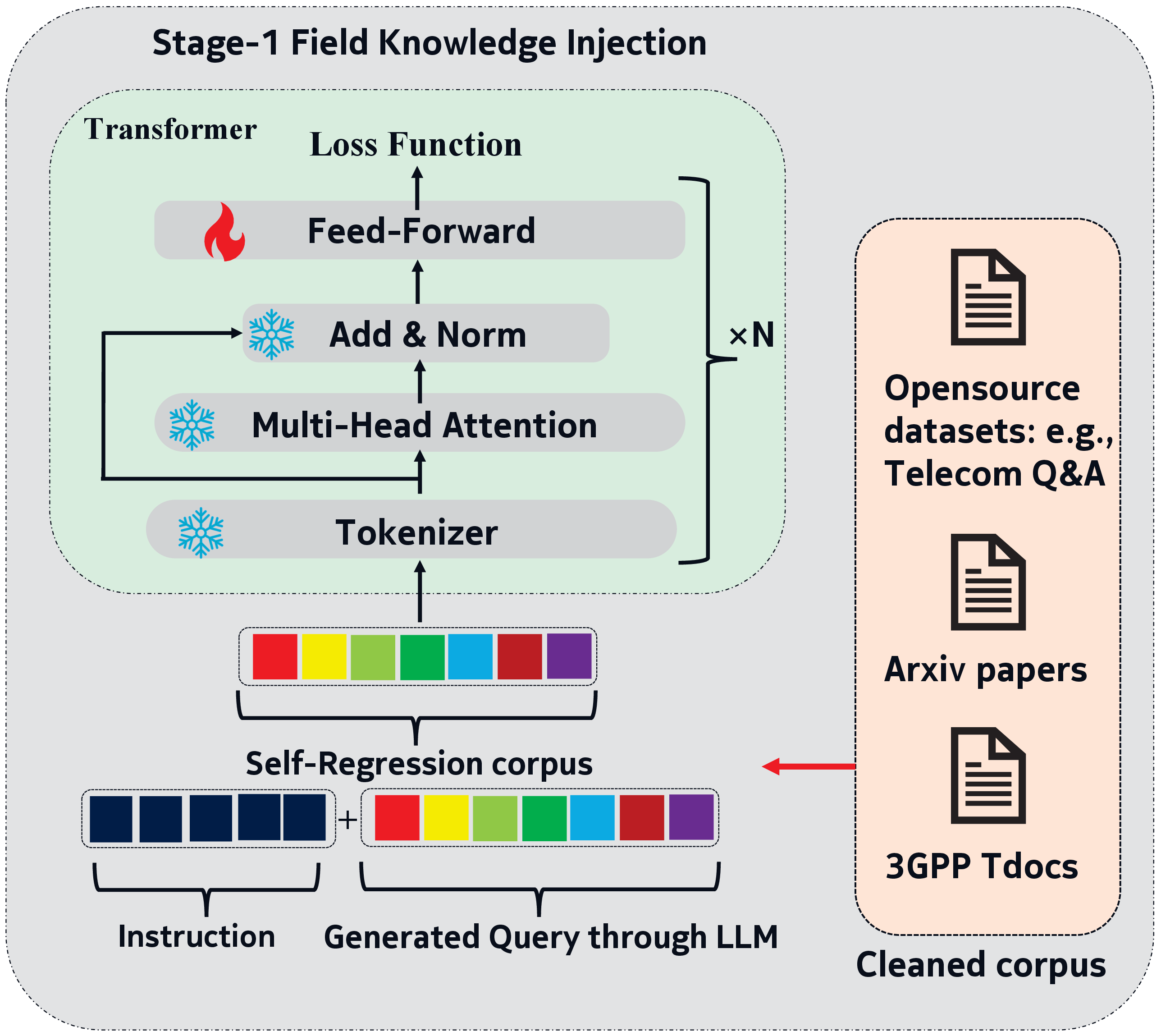}
		\end{minipage}

	}
	\subfigure[Stage 2: LoRA Fine-Tuning] {\label{lora}
		\begin{minipage}[t]{0.7\linewidth}
			\centering
			\includegraphics[width=1\textwidth]{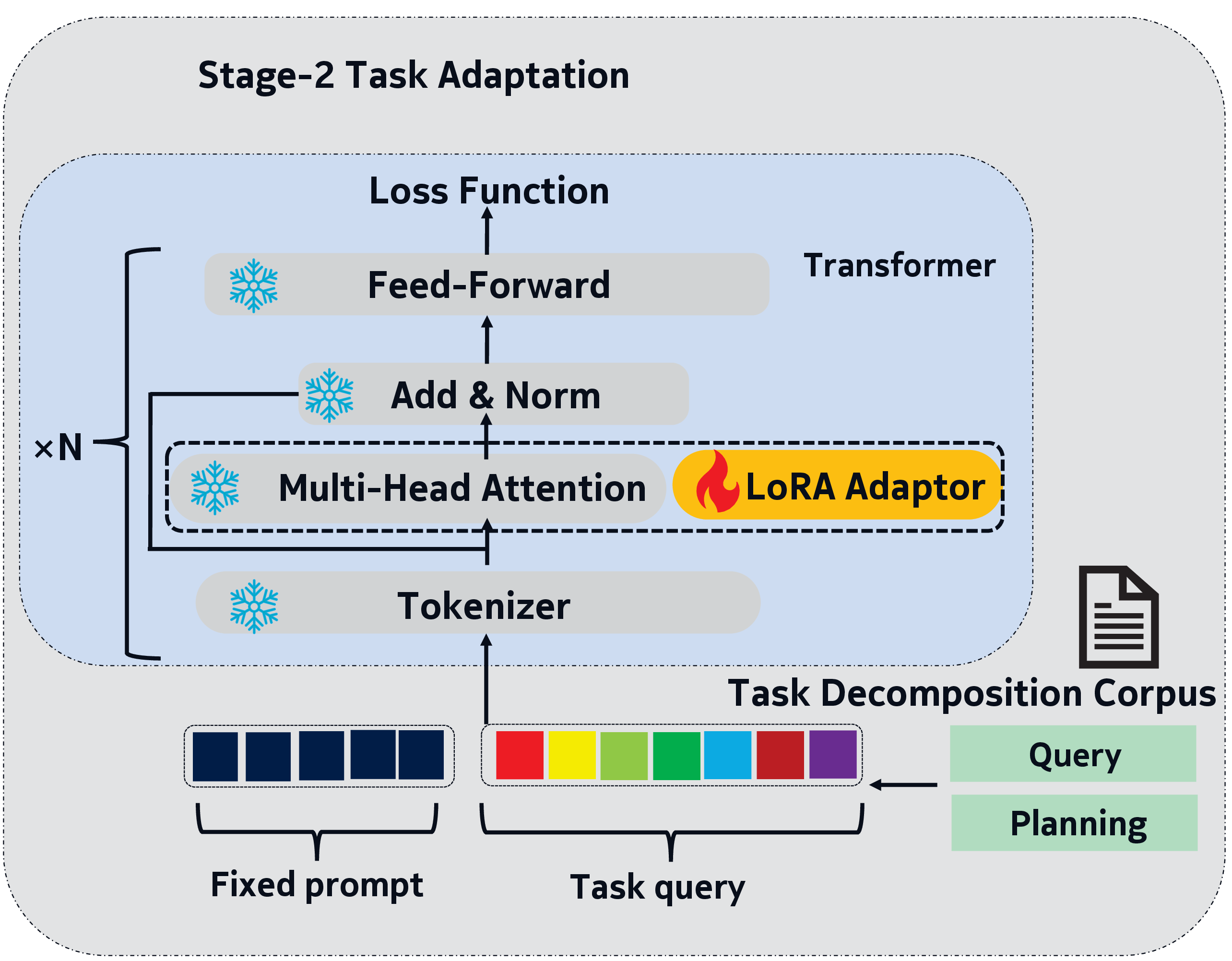}
		\end{minipage}
	}
	\centering
	\caption{Learning structure of two-stage training process.}
	\label{nmse2}
	\vspace{-.5cm}
\end{figure}

\begin{figure*}
	\centering
	\includegraphics[width=0.85\textwidth]{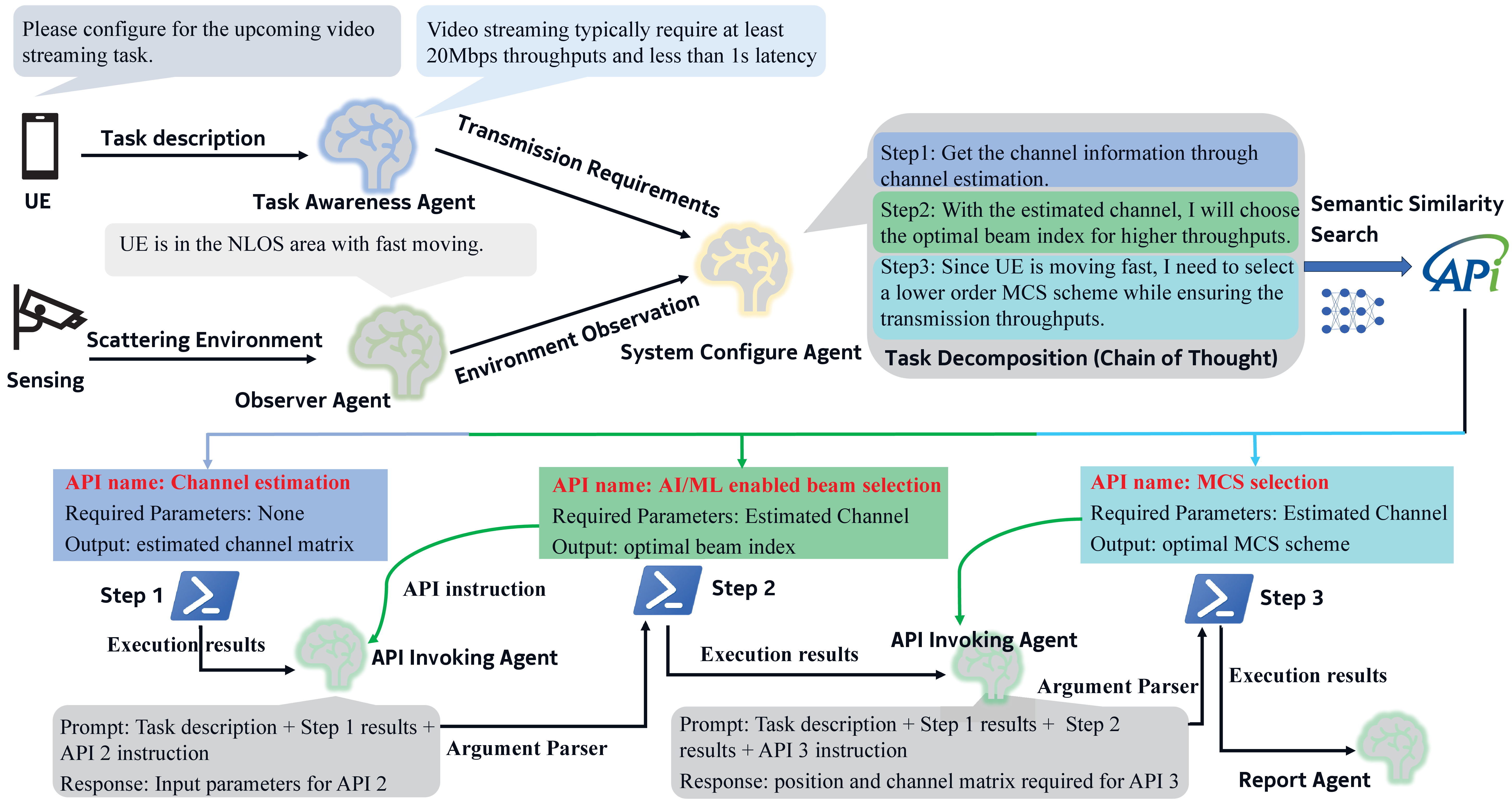}
	\caption{Example of the workflow of a communication system orchestrated by LLM agents. LLM agents play different roles in different parts of the system. The agents could understand UE's upper-layer task requirements and orchestrate the system step by step in an optimal way accordingly.}
	\vspace{-.5cm}
	\label{workflow}
\end{figure*}
\subsection{Stage One: Knowledge Injection Through Continual Learning}
In the stage-1 training process, we propose a continual training scheme by leveraging a pre-training general-purpose LLM. The rationale behind utilizing a pre-training model lies in its capacity to leverage the inherent understanding of language directly. Further, we propose to train only the parameters within the fully connected layers of the pre-training model's top $K$ transformer blocks rather than engaging in full parameter training. This idea is based on several key considerations. Firstly, despite the massive scale of available training datasets, it remains insufficient to effectively train a complete model with billions of parameters without causing significant overfitting issues. Secondly, catastrophic forgetting is a prevalent challenge when training large language models. This effect is pronounced when training with highly structured and simplified corpus, which is common in the telecommunication domain, such as the protocol documents and function instructions. The attention layer in the LLM works to maintain the implicit token correlations at the natural language level, which would be easily ruined by those data. Thus, by selectively training only the uppermost feed-forward layers while freezing the underlying attention layers, we significantly enhance the robustness and mitigate the risk of catastrophic forgetting in the LLM.

The training task is set to a combination of unsupervised self-regression and supervised instruction following. The unsupervised self-regression approach is widely employed in the training regimen of LLMs and eliminates the need for labeled data. However, this learning methodology can lead to potential inefficiencies in the learning process and difficulty in directing the model's attention toward crucial information. Thus, we propose to involve the utilization of a generic LLM to extract the critical information of the prompted document and generate high-quality labeled data in the format of question-answer pairs. This way, the model pays more attention to information of higher importance, such as technical terminology. This way, we greatly enhance the model's expert knowledge while maintaining robustness.

\subsection{Stage Two: Role-Specific Fine-Tuning}
The objective of the stage-2 training regimen is to refine the model's propensity for making decisions tailored to the specific role and function of the target agent. For example, an environment observing agent should typically exist that keeps monitoring the changing of the transmission environment and a system-configure agent that changes the system configuration accordingly. A distinguishing attribute among different agents lies in their respective API databases. Therefore, the initial step of data preprocessing of this step is the creation of comprehensive API database documentation, which encapsulates meticulous descriptions of the functionalities, parameter definitions, and output specifications. Subsequently, this enables the procurement of a query and task decomposition corpus. Queries within this corpus can manifest in diverse formats, from direct task descriptions to specific transmission requirements. Every composite step in the corpus corresponds to a potential API within the established database.

We propose to adopt the LoRA method to fine-tune the L1 model with a supervised fine-tuning scheme. This involves appending a low-rank adapter to each parameter matrix within the original transformer architecture, thereby enabling the model to swiftly adapt its planning and decision-making capabilities according to its unique tool package. Adopting this training configuration presents several salient benefits. Primarily, the compact size of the LoRA adapters allows them to be effortlessly updated in the cloud server and subsequently downloaded by the Base Station (BS) or User Equipment (UE), thus guaranteeing operational flexibility. Furthermore, the LoRA technique exhibits a lower propensity for causing catastrophic forgetting or overfitting when dealing with datasets of limited size, thereby ensuring the overall robustness of the entire system. It is worth mentioning that each query in the training corpus is concatenated with a fixed prompt that describes the agent's role, which will help enhance the convergence of the training process.

\section{Workflow Design for Task-Oriented Physical-Layer Automation} \label{agent}
\subsection{Role-Playing and Collaboration of Agents}
There are two main concerns regarding clearly defining LLM agents' different roles in the system. Firstly, from the point of practical usage of LLM, clear role definitions facilitate the optimization of prompt design, thereby enhancing the accuracy of model outputs and mitigating the likelihood of model hallucinations. Secondly, it will help facilitate streamlining system processes and enhance system efficiency since each role has straightforward triggering logic.

In the proposed system workflow shown in Fig. \ref{workflow}, we define four LLM agent roles to finish the system workflow. For each coming UE task, the task awareness agent will first translate the upper-layer task description to the precise transmission requirements. The observer agent will constantly monitor the scattering environment and summarize the system state. Each time there is a new UE task or changing environment, the system configuration agent will give its reasoning and planning with the complete chain of thoughts. After API retrieval, the API invoking agent will help run those APIs correctly. After all the abovementioned processes, a reporting agent will summarize the executing results for potential error-catching and shooting.

\subsection{API Retrieve Through Vector-Based Semantic Similarity Search}
In existing works, tool utilization relies on putting all the API descriptions into the prompts. This approach proves practical for simple tasks when the number of APIs involved is limited. However, there are thousands of APIs for the complex communication system, and it is not practical to include all in the prompt due to the context length constraints. To address this issue, we propose a novel solution by semantic vectorizing API instructions in an extended vector database and utilizing similarity search for retrieval.

Firstly, we use a pre-trained semantic sentence-to-vector model to encode all the API instructions into fixed-length vectors. Then, for each output step of the LLM, we apply the same model to generate a semantic representation vector. Then, we calculate the GCS (Generalized Cosine Similarity) between the generated step vector and each API embedding stored in the database, which can be written as
\begin{equation}
GCS=\frac{\mathbf{D}\cdot \mathbf{S}}{\left \| \mathbf{D} \right \| \cdot \left \| \mathbf{S} \right \|},
\end{equation}
where $\mathbf{D}$ is the embedding vector of the items in the API database and $\mathbf{S}$ denotes the embedding vector of the planning steps by the LLM. The API with the highest similarity score is then selected as the most relevant match. This strategy allows for efficient and scalable API retrieval even when the API inventory is vast and diverse.

\subsection{Tool Execution Based on Self-Organization}
How to accurately plan the invocation of tools and successfully execute them by finding the appropriate actual parameters for each API has always been a challenging issue. In existing works, the common approach is rigorously constraining the parameter format of the APIs and implementing argument filtering in the program. Thus, all the tools should be developed specifically at the beginning stage, or extra efforts should be made to refine them to meet the format requirement. However, there exists a huge library of API functions developed for different protocol standards, which are impractical to switch to a common format. Therefore, we propose employing LLMs as an intermediary to bridge these diverse APIs and effectively identify the requisite parameters.

As shown in Fig. \ref{workflow}, it is illustrated that the API invoking agent will step in every time prior to executing an API and find the corresponding input parameters. The parameter inputs have three sources: values in the task descriptions, outcomes from previously executed APIs, and default values inherent to the current API under execution. Thus, we integrate all the sources in the prompt and require the LLM to find the value through physical meaning automatically. For example, if the former API instruction mentions that the estimated channel state information matrix will be stored in a temporary variable named `\textit{CSI\_matrix}' and the under-executing API's instruction mentions that it requires the estimated CSI as the value of a parameter named `\textit{estimated\_channel}', the LLM will automatically realize that `\textit{estimated\_channel} = \textit{CSI\_matrix}'. It is worth mentioning that this method is not trivial since it requires the LLM to recognize those terminologies and connect them with the descriptions, which is highly reliant on expert knowledge injection. Thus, the API-invoking agent needs to be driven by our L1 model, which ensures the smoothness and continuity of system actions.

\section{Experiments} \label{experiment}
\subsection{Parameters Setting and Datasets Generation}
We employ two different sizes of open-source LLM models, specifically LLaMA2-13B and LLaMA2-70B, as our foundation pre-trained L0 models. During the stage-one continual training, we train the parameters within the feed-forward layers of the uppermost transformer blocks, and the total trainable parameters are approximately 3.5 billion for both the 13B and 70B models. In stage-two training, we further adopt the LoRA method for all matrices within the attention blocks (Query, Key, and Value) across the entire transformer network. The adapter rank is set to $16$, thereby introducing approximately an additional $0.6\%$ of trainable parameters compared with the original model.

%
%
%
%
%
%
%
%
%
%
%
%
%

\subsubsection{Dataset Generation}
For the stage one training, the datasets encompass diverse textual resources, including 3GPP documentation, ArXiv papers related to telecommunication, and open-source datasets such as \textit{Telecom Q\&A} \cite{10437725}. To ensure the quality of the datasets, all original texts undergo rigorous preprocessing procedures involving comprehensive data cleaning and structural formatting. Additionally, we employ Tongyi-Max to help clean and rephrase some corpus. Thus, the information-intensive parts will be highlighted.

For stage two fine-tuning, we initially constructed a tool library comprising approximately $200$ APIs. Subsequently, we manually create about $20000$ in corpus composed of UE tasks, transmission requirements, and corresponding decomposited action steps. Each action step corresponds to an existing API in the tool library.

\subsection{Experimental Results}
\begin{figure}[b!]
		\vspace{-.5cm}
	\centering
	\includegraphics[width=0.35\textwidth]{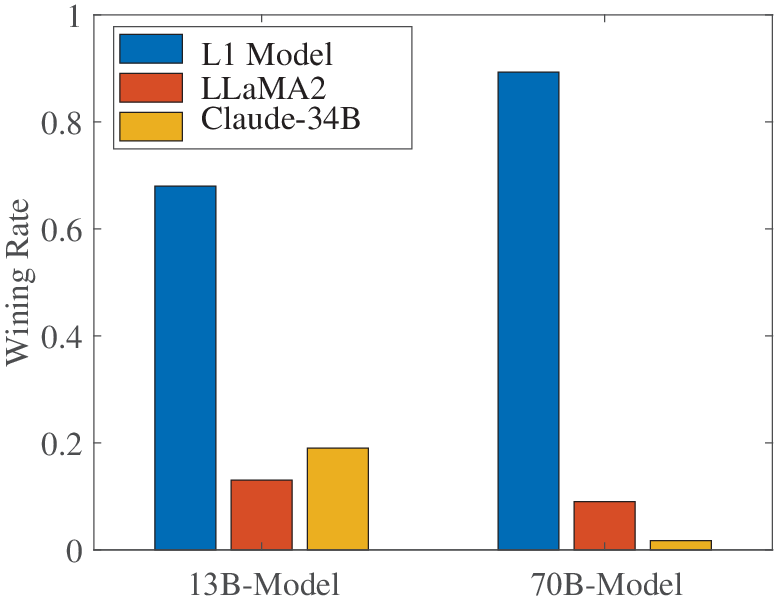}
	\caption{The winning ratio comparison of answering communication related questions between three LLMs.}
	\label{exp1}
\end{figure}

\begin{figure}[b!]
	\centering
	\includegraphics[width=0.35\textwidth]{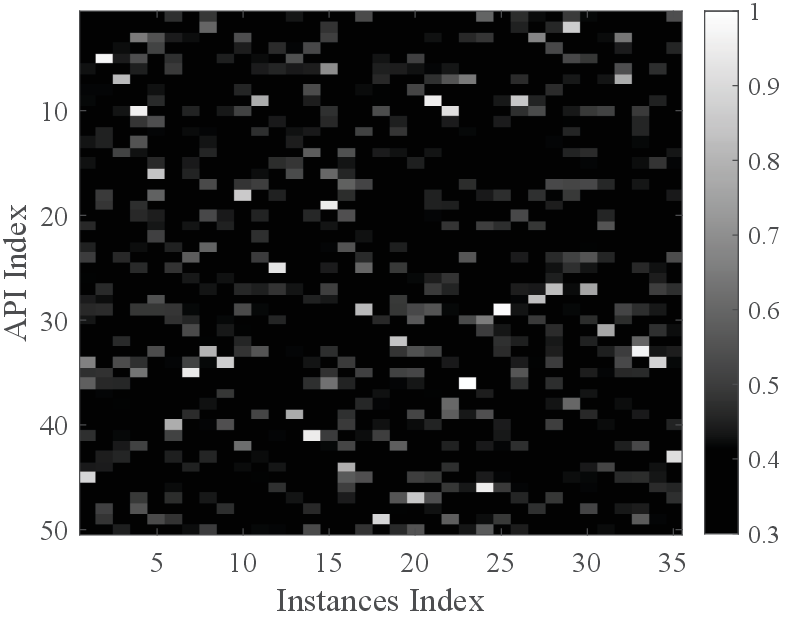}
	\caption{The cosine similarity measurement is computed between the embedding vector representative of the $35$ exemplary step descriptions and the respective embedding vectors of the API instructions within the database.}
	\label{exp3}
\end{figure}

\begin{figure}
	\centering
	\includegraphics[width=0.35\textwidth]{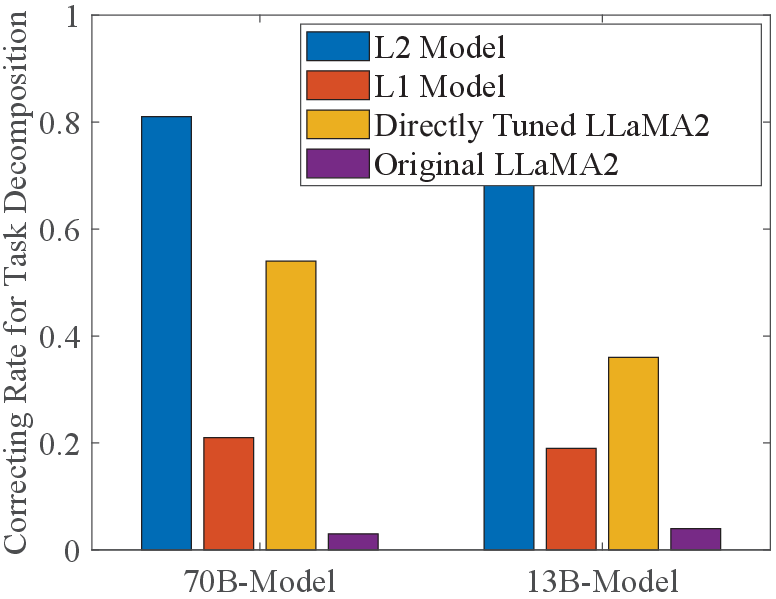}
	\caption{The correction rate comparison of task decomposition between the comparison models.}
	\label{exp2}
\end{figure}

\begin{figure}
	\centering
	\includegraphics[width=0.5\textwidth]{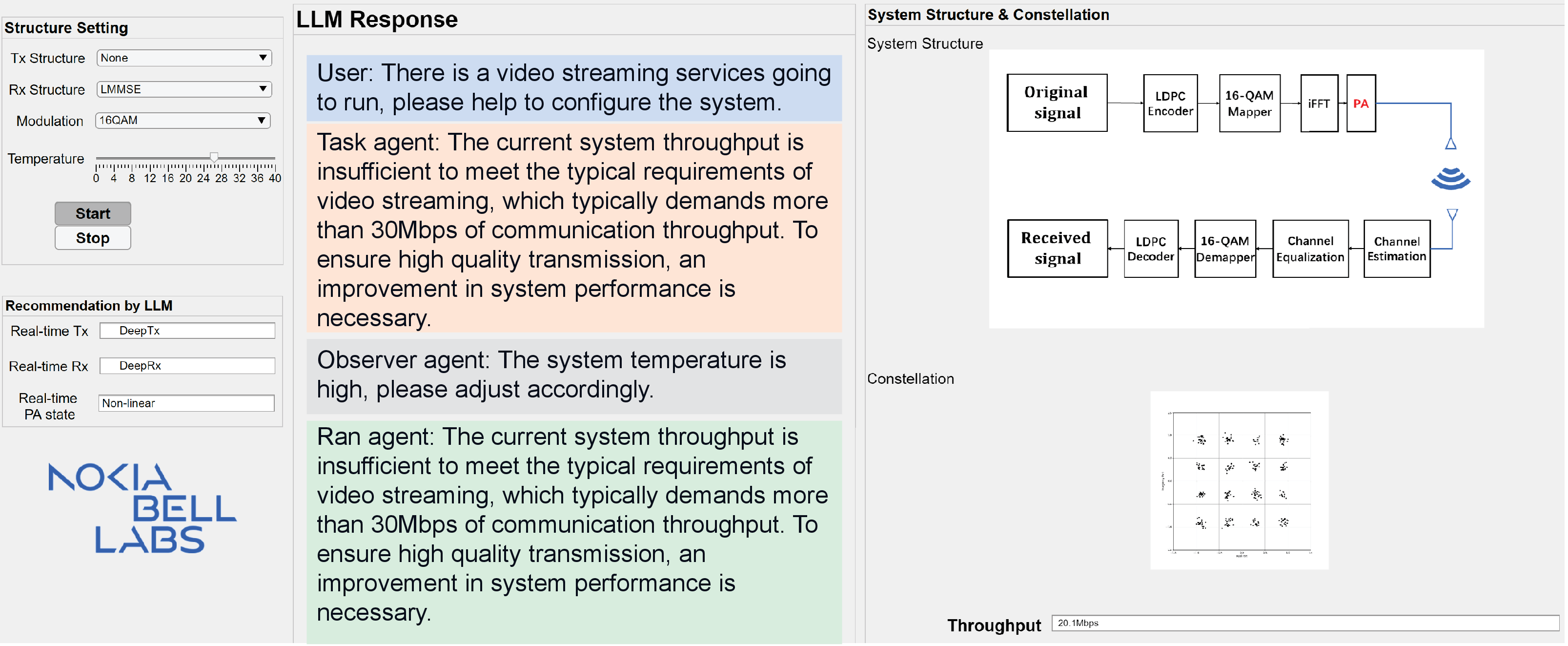}
	\caption{A conceptual demo of adopting LLM agents to orchestrate a practical communication system.}
	\label{deep_demo}
\end{figure}

Firstly, we compare the performance of our L1 model against the LLaMA2 and Claude when answering telecommunication-related questions, thereby verifying the effect of domain-specific knowledge injection. We gather a testing dataset consisting of $4000$ 3GPP protocol-related questions, accompanied by their correct answers and corresponding explanations. These questions are presented to three comparative models with the same prompt. A third-party LLM, Tongyi-max, is prompted with all the questions alongside their respective answers and explanations. The task assigned to Tongyi-max is to evaluate which of the provided answers aligns more closely with the correct answer. Fig .\ref{exp1} shows the overall winning ratio for the three models. Evidently, with the implementation of knowledge injection, the L1 model consistently outperformed its counterparts across both the 13B and 70B models, proving the effectiveness of training.

Secondly, to illustrate the feasibility of the proposed semantic similarity search-based API retrieval method. We randomly selected $35$ exemplary step instances outputted by the L2 model, which should correspond to the first $50$ APIs in the library and obtain the semantic embedding vectors. Then, we calculate the GCS between the step description embedding vectors and the API instruction embedding vectors. Results are shown in Fig. \ref{exp3}. It can be observed from the results that for each instance, the correct API always has the highest similarity, significantly surpassing other wrong APIs, which ensures that we can always find the correct API for each planning step.

Thirdly, to further validate the effectiveness of the stage-two training, we collect a validation dataset comprising $1000$ query instances across various styles, along with their corresponding optimal decomposition steps. Then, we compare the response of the L2 model with three comparison models shown in Fig. \ref{exp2}, which include the original LLaMA2 model, the L1 model, and the directly fine-tuned model without knowledge injection continual training. To ensure a fair comparison, we employed the identical prompt and used the above-mentioned method to call the corresponding API of each step and compare the results with the grand truth. Thus, we can calculate the correcting rate of each model. As shown in Fig. \ref{exp2}, the proposed L2 model exhibits the highest accuracy ratio and greatly surpasses other components, which proves the effectiveness of the stage-two training process. Besides, a noteworthy observation is that, following stage-one training, the L2 model significantly outperforms the model that was directly fine-tuned without any knowledge injection. This result further highlights the feasibility of our proposed training methodology.

Finally, in order to intuitively illustrate the advantages of applying LLMs in wireless communication systems, we have made a conceptual demo of adopting LLM agents for automatically orchestrating the invocation of DeepTx and DeepRx \cite{9345504} to enhance system throughput and address PA nonlinearity issues caused by rising device temperatures. The detailed video demo can be obtained in `\textit{https://github.com/hongan-nokia/bell\_labs\_6G\_llm}' for further reference.

\section{Conclusions} \label{conclusion}
This paper proposed a novel paradigm for training and applying LLM agents for task-oriented 6G physic-layer automation.
By employing a specifically designed two-stage continual training approach, the trained domain-adapted LLM effectively aids in understanding the upper-level requirements of users and accordingly recommends the workflow and the optimal system configurations.
Further, with the proposed workflow of building the automation system, the responses of those LLM agents can be mapped to the actual invoking of corresponding APIs, which finally orchestrate the system for achieving higher performance. Experimental results validated the feasibility and effectiveness of the proposed methodology and its potential to tackle practical challenges in the future design of wireless communication systems.

\bibliographystyle{IEEEtran}
\bibliography{bibfile}
\end{document}